\def \SAIT #1 #2 {{\em Mem.\ Soc.\ Astron.\ It.\/} {\bf #1}, #2}
\def \MESS #1 #2 {{\em The Messenger\/} {\bf #1}, #2}
\def \ASTRNACH #1 #2 {{\em Astron. Nach.\/} {\bf #1}, #2}
\def \AAP #1 #2 {{\em Astron. Astrophys.\/} {\bf #1}, #2}
\def \AAL #1 #2 {{\em Astron. Astrophys. Lett.\/} {\bf #1}, L#2}
\def \AAR #1 #2 {{\em Astron. Astrophys. Rev.\/} {\bf #1}, #2}
\def \AAS #1 #2 {{\em Astron. Astrophys. Suppl. Ser.\/} {\bf #1}, #2}
\def \AJ #1 #2 {{\em Astron. J.\/} {\bf #1}, #2}
\def \ANNREV #1 #2 {{\em Ann. Rev. Astron. Astrophys.\/} {\bf #1}, #2}
\def \APJ #1 #2 {{\em Astrophys. J.\/} {\bf #1}, #2}
\def \APJL #1 #2 {{\em Astrophys. J. Lett.\/} {\bf #1}, L#2}
\def \APJS #1 #2 {{\em Astrophys. J. Suppl.\/} {\bf #1}, #2}
\def \APSS #1 #2 {{\em Astrophys. Space Sci.\/} {\bf #1}, #2}
\def \ASR #1 #2 {{\em Adv. Space Res.\/} {\bf #1}, #2}
\def \BAIC #1 #2 {{\em Bull. Astron. Inst. Czechosl.\/} {\bf #1}, #2}
\def \JSQRT #1 #2 {{\em J. Quant. Spectrosc. Radiat. Transfer\/} {\bf #1}, #2}
\def \MN #1 #2 {{\em Mon. Not. R. Astr. Soc.\/} {\bf #1}, #2}
\def \MEM #1 #2 {{\em Mem. R. Astr. Soc.\/} {\bf #1}, #2}
\def \PLR #1 #2 {{\em Phys. Lett. Rev.\/} {\bf #1}, #2}
\def \PASJ #1 #2 {{\em Publ. Astron. Soc. Japan\/} {\bf #1}, #2}
\def \PASP #1 #2 {{\em Publ. Astr. Soc. Pacific\/} {\bf #1}, #2}
\def \NAT #1 #2 {{\em Nature\/} {\bf #1}, #2}
\documentstyle[psfig]{memsait}
\input epsf.sty

\begin{opening}
\title{The AGILE gamma-ray detector  $^{(\star)}$ }
\author{A. Morselli$^1$, A. Perrino$^1$, P. Picozza$^1$ S. Severoni$^1$ , P.
Caraveo$^2$, S.~Mereghetti$^2$, M. Tavani$^{2,4}$,  G.Barbiellini$^3$, 
 A. Vacchi$^3$  }
\institute{ $1$. Department of Physics, University of Rome 2 ''Tor Vergata''
and INFN Rome 2, Italy\\
$2$. Istituto di Fisica Cosmica G.Occhialini, CNR, Milano\\
$3$.  Department of Physics, University of Trieste and INFN Trieste\\
$4$. Columbia Astrophysics Laboratory, Columbia University, New York, USA}
\date{} 
\end{opening}
\begin{document}
\vskip -1.cm
 
\oddpagefooter{}{}{} 
\evenpagefooter{}{}{} 
\ 

\begin{abstract}

\noindent The gamma-ray detector AGILE, operating in the energy range from 30
MeV to 50 GeV,  is composed by a tracking part, a light calorimeter
and an anticoincidence system.  Here we describe the detector and its 
capabilities to determine the    arrival direction and   energy of the
detected photons.

\end{abstract}

\section{\bf The AGILE detector }

The AGILE design is derived from a refined study of the GILDA project
(Barbiellini et al. 1995, Morselli et al. 1995)
that  was based on the techniques developed for the Wizard
silicon calorimeter,  already successful flown in balloon
experiments (Golden et al. 1990).
 Like previous orbiting high-energy gamma-ray telescopes, AGILE relies on the
unambiguous identification of the incident gamma-rays by recording the
characteristic track signature of the e$^+$ - e$^-$   that result from pair
creation from the incident photons in thin layers of  converter material. The
trajectories of the  pairs,   recorded and measured in the tracking section of
the detector,  give the information on the direction of the incident
gamma-rays.  A light calorimeter made of 1.5 radiation lengths (X$_0$) of
cesium iodide (CsI)  allows to determine   the
energy of the incident photons.  The detector has a height of $\sim$35 cm, with a $40 \times 40$ cm$^2$
area and a total conversion length, for electromagnetic cascades, of $\sim$2.3
X$_0$.  A schematic view of the AGILE detector  is presented
in Tavani et al. (1998).
 The three AGILE components are:
{\sl i}) a {\bf  tracker}: this is the conversion zone. It consists of 12
planes of single sided silicon strips with 204 $\mu$m pitch 
(distance between strips). The strips are implanted on
pads of $8\times 8$ cm$^2$. Each pad has a  thickness of 380 $\mu$m and 
carries 384 strips.  The first ten planes consist of two layers of silicon detectors,
with orthogonal  strips interspersed with a layer of 0.07 X$_0$ of tungsten. 
The last two planes have no tungsten sheets. Each plane reads 3840 strips and
the entire tracker consists of 46080 strips.  The distance between consecutive 
planes is 1.6 cm;
{\sl ii}) a {\bf  light calorimeter}: it is made of 1.5 X$_0$ of CsI and will
provide an indication of the photon energy;
{\sl iii}) an{ \bf  anticoincidence system}: it consists of 0.5 cm thick
plastic scintillator layers surrounding the top and four lateral sides of the
silicon tracker. It will be able to
significantly reduce the background due to charged particles.

\noindent
------------------------------------------------------------------------------- \\
($\star$) Adapted from a paper presented at the Conference
{\it Dal nano- al Tera-eV: tutti i colori degli AGN}, Rome 18-21 May 1998,
to be published by the {\sl Memorie della Societa' Astronomica Italiana}.
\newpage
  
The expected performance of AGILE in angular and energy resolution,
as derived from Montecarlo simulations,
  are shown in Fig.~1. 
The reliability of
the Montecarlo simulations, based on the GEANT code, has been checked with 
experimental data obtained during tests performed at CERN 
(Bocciolini et al. 1993). 

Despite its smaller area and weight, the AGILE
instrument is able to reach   performances comparable to those of   the
EGRET experiment on   CGRO.
The main
characteristics that make this telescope more efficient are:
 {\sl i}) the use of silicon strips instead of a spark chamber  as the  main device
for reconstructing  the gamma-ray  trajectories;
this also  avoids the problems of gas refilling, high voltages and 
dead time (silicon strips have lower dead time than gas detectors)
and allows to    reach a spatial
resolution of two hundred  microns, instead of a few millimeters.
{\sl ii}) the elimination of the anticoincidence counters for the high energy
trigger, so that 
 an efficiency of 50\% up to 50 GeV can be reached; 
 {\sl iii}) the elimination of Time of Flight, resulting in an increase
of the field of view and a decrease of the lower  energy threshold 
($\sim$20 MeV for AGILE vs. 35 MeV for EGRET).
The impact of theese performances on the study of the AGN's is described in 
Mereghetti et al. (1998).

\begin{figure}[ht]
\vspace*{-0.5cm}
\psfig{figure=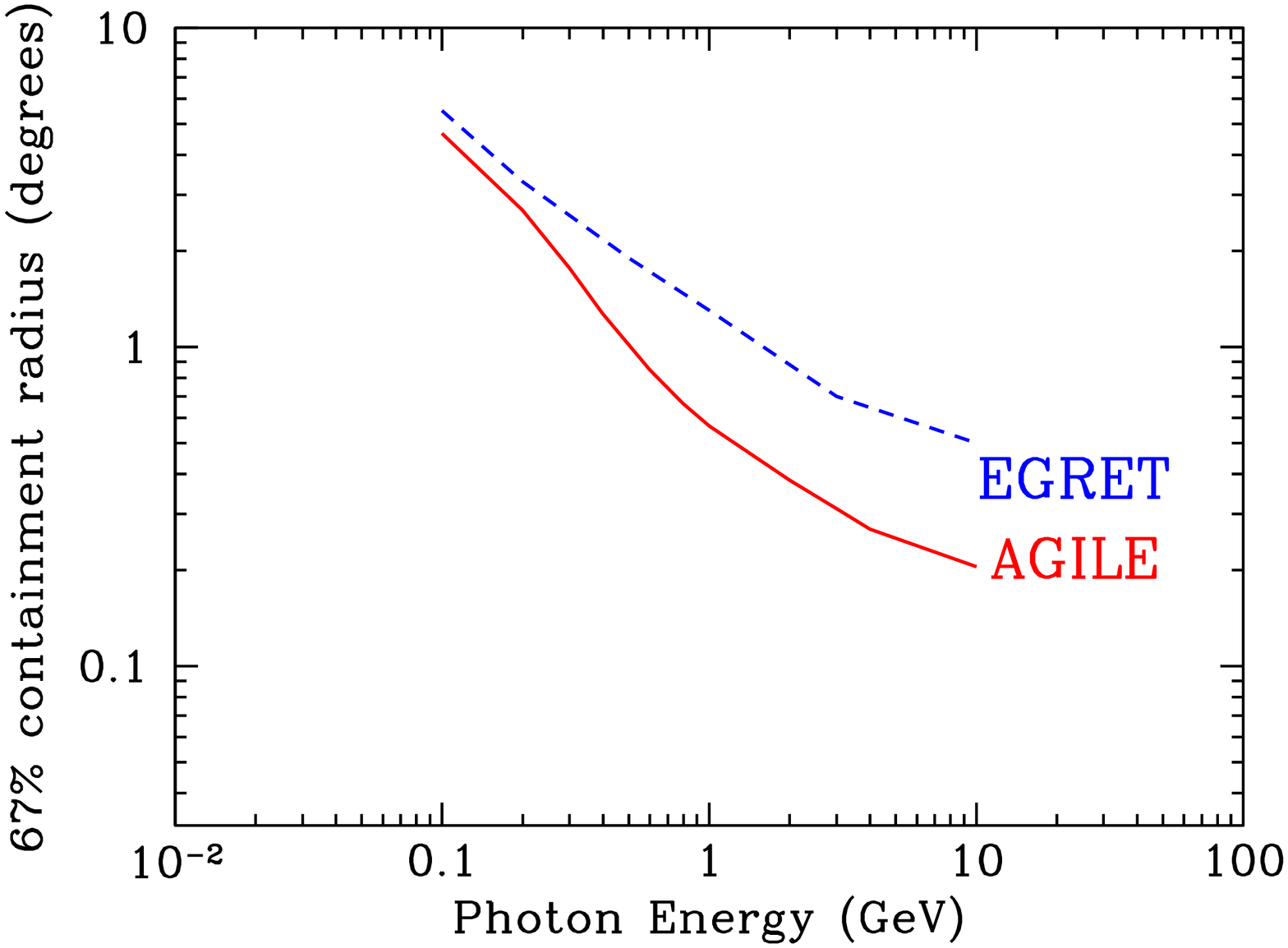,width=6cm,height=6cm,angle=-90}
\vspace*{-6cm}
\hspace*{7cm}
\psfig{figure=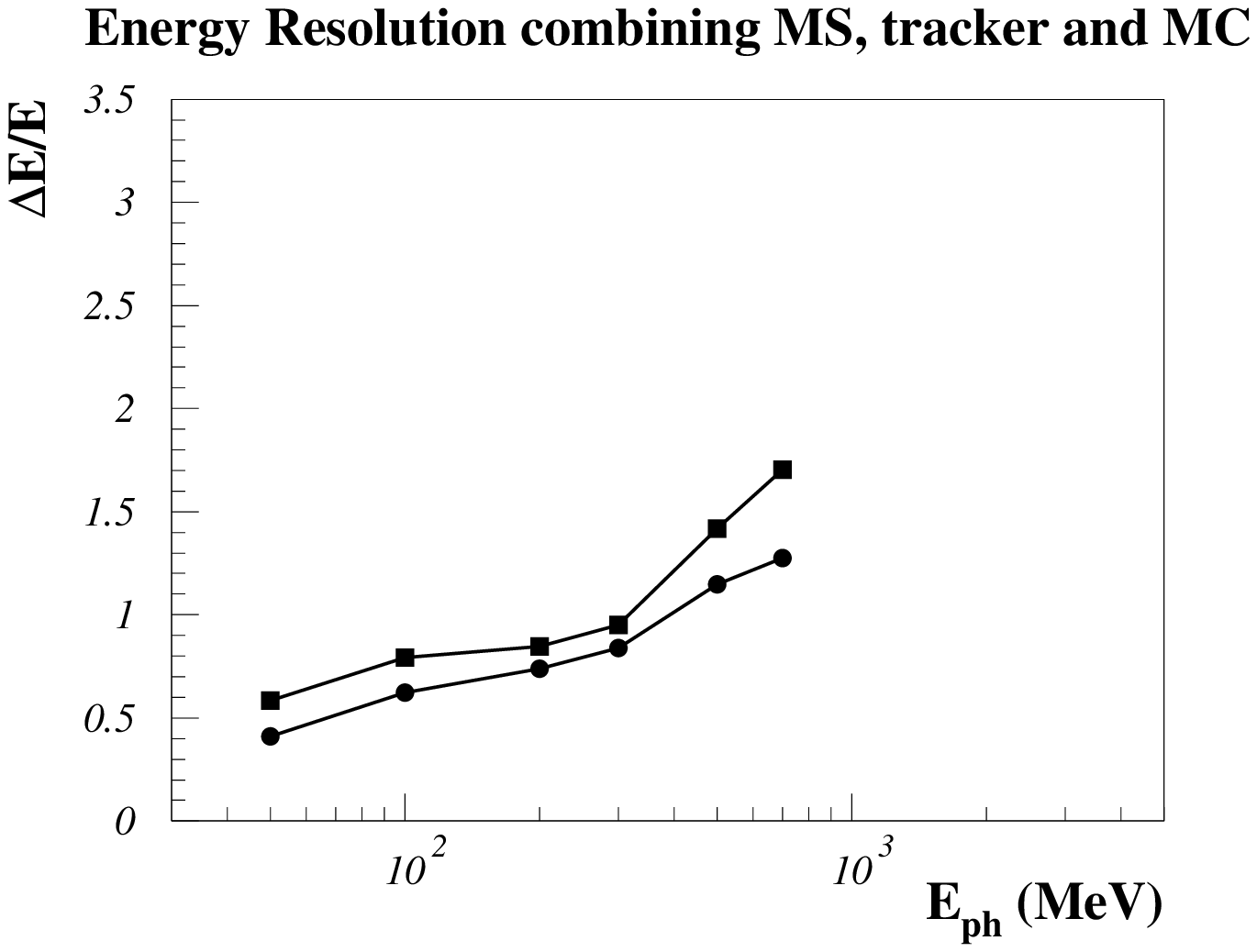,width=6cm,height=6cm,angle=0}
\vspace*{-0.5cm}
\caption{Three dimensional PSF (67\% containment radius)
as a function of photon energy  for AGILE and EGRET (on the left). 
AGILE expected energy resolution (on the right) obtained combining
the multiple scattering method (MS), the energy released into the 
mini-calorimeter (MC) and into the tracker for a photon converted in the 
first plane ({\it circles}) and in the last plane ({\it squares}).}
\end{figure}


\begin{thebibliography}{}  

\bibitem 
[ Barbiellini G. et al. 1995]{gilda$nim}   Barbiellini G.  et al.: 1995,
{\sl NIM}, {\bf A354}, 547. 

\bibitem [ M. Bocciolini et al. 1993]{wizsim}  Bocciolini M. et al.: Nuclear
Instruments and Methods , {\bf A333}, 560.


\bibitem 
[ Golden  R.L. et al. 1990]{Wizard} Golden R.L. et al.:1990, {\sl Il
Nuovo Cimento}, {\bf 105 B}, 191.

\bibitem [ S.Mereghetti et al. 1998] {agiles}  Mereghetti S. et al., 1998:
theese proceedings. (astro-ph/9811444)

\bibitem
[Morselli A. et al. 1995]{rim}     Morselli A. et al.: 1995,  XXIV
ICRC, OG 10.3.26, v.3, 669, Roma.

\bibitem 
[Tavani M. et al. 1998] {agilet}  Tavani M. et al., 1998: theese
proceedings. (astro-ph/9812096)


\end{thebibliography}
\end{document}